\documentclass[conference]{IEEEtran}
\usepackage[T1]{fontenc}
\usepackage[latin9]{inputenc}
\usepackage{color}
\usepackage{amssymb}
\usepackage{graphicx}
\usepackage{esint}
\usepackage[bookmarks=true,bookmarksnumbered=true,bookmarksopen=true,bookmarksopenlevel=1,
 breaklinks=false,pdfborder={0 0 0},pdfborderstyle={},backref=false,colorlinks=false]
 {hyperref}
\hypersetup{pdftitle={Your Title},
 pdfauthor={Your Name},
 pdfpagelayout=OneColumn, pdfnewwindow=true, pdfstartview=XYZ, plainpages=false}

\makeatletter
\usepackage[caption=false,font=footnotesize]{subfig}

\makeatother

\begin{document}
\title{Pink noise in Electric Current from Amplitude Modulations}
\author{\IEEEauthorblockN{Masahiro Morikawa}\IEEEauthorblockA{Department of Physics, Ochanomizu University\\
2-1-1 Otsuka, Bunkyo, Tokyo 112-8610, Japan\\
hiro@phys.ocha.ac.jp}\and \IEEEauthorblockN{Akika Nakamichi}\IEEEauthorblockA{General Education, Kyoto Sangyo University\\
Motoyama Kamigamo Kita-ku, Kyoto 603-8555, Japan\\
nakamichi@cc.kyoto-su.ac.jp}}
\IEEEaftertitletext{}
\maketitle
\begin{abstract}
We recently proposed that the general origin of 1/f fluctuation, or
pink noise, is the amplitude modulation (or beat) of many waves with
accumulating frequencies. In this paper, we verify this proposal in
the electric current system. We use the classical Langevin equation
to describe the electron wave packets flowing in the (semi-)conductor,
affected by the back reaction of soft photon emission. If this were
amplitude modulation, a demodulation process is needed to extract
the fluctuation features. We first square the wave packet, which corresponds
to the electric current, and obtain the 1/f fluctuation in this current
data. We further speculate that this wave packet, after demodulation
by thresholding, triggers the time sequence of the nerve firing. In
our model, this also shows 1/f fluctuations, which is quite robust. 
\end{abstract}

\IEEEpeerreviewmaketitle{}

\section{Introduction}

1/f fluctuation, or pink noise, appears everywhere in nature. The
pink noise is characterized, in the power spectrum density (PSD),
by the low-frequency power behavior with an index $-1\pm0.5$. Pink
noise is observed in semiconductors, thin metals, potential fluctuations
in bio-membranes, electric currents, long-term temperature fluctuations,
flow fluctuations on highways, music, variations in the intensity
of cosmic rays, heartbeat rate, MEG, and EEG (brain), etc.\cite{Johnson1925,Milotti2002,Bosnian2001,Hooge1994}.
\quad{}

 Many theories have been proposed so far to reveal the origin of
pink noise\cite{Milotti2002}. But we have no decisive theory yet.
Some representative theories are flicker noise theory, surface defect
scattering theory, two-level systems theory, charge trapping and de-trapping
theory, fractal geometry theory, burst noise theory, self-organized
criticality theory, power-law tunneling model, fractional calculus
theory, and many more. However, none of them are universal explanations. 

We realize that the authors of the proposed theories seem to seek
real fundamental fluctuations. Since pink noise is common, it may
not have any complex deep origin but some \textbf{common interference
effect} \cite{Morikawa2021}. Further, we recognize a curious fact
that many authors have needed to apply the square of the data before
Fourier analysis. Therefore, pink noise may not appear in the original
signal. We expect this square operation may represent some\textbf{
demodulation} of the encoded pink noise data. All these considerations
lead to our proposal that the pink noise originates from amplitude
modulation or wave beat \cite{Morikawa2023}. This is explained in
section II. 

Regarding pink noise in the electric current, Handel proposed the
quantum origin: incoming electron wave function and the scattered
electron wave function, after emitting soft photons, interfere with
each other to yield pink nise\cite{Handel1975,Handel1980}. However,
these two states turn out to be orthogonal since the photon number
is different in the two states\cite{Kiss1986,Nieuwenhuizen1987}.
Therefore, we learn that the pink noise in the electric current is
classical interference. This consideration leads to the wavepacket
dynamics, described in section III.

We further expect that the pink noise in the electric current may
be inherited by the nerve pulse through thresholding\cite{Stern1997}.
If the pink noise is robust, the living body may inherit the fluctuation
property. This possibility is described in section IV. 

We describe possible verification/falsification methods of our proposal
in section V. The final section VI is the conclusions of the present
work and prospects for future work. 

\section{Amplitude modulation - the origin of pink noise}

We have recently proposed a simple physics for the origin of pink
noise\cite{Morikawa2023}. Pink noise that characterizes the low-frequency
region generally comes from the wave beat of accumulating frequencies.
This is amplitude modulation. If the pink noise comes from the wave
beat, it cannot appear in the PSD of the original signal of the superposed
many waves. However, pink noise appears in the PSD of the squared
signal. 

This is fully consistent with many previous features of pink noise.
For example, in music \cite{Voss1977}, the square operation of the
original sound data was needed to yield pink noise. This is interpreted
that the pink noise appears in the loudness and not in the displacement
itself. Also, in the Hamiltonian Mean Field (HMF) models\cite{Yamaguchi2018},
the calculated mean field $M$ itself does not show pink noise, but
$M^{2}$ does. Further, in the original Johnson's work on the current
fluctuations in the vacuum tube, voltage fluctuation squared is needed
to show pink noise. 

The frequency accumulation property determines the power index\cite{Morikawa2023}.
We have already proposed typical mechanisms of frequency accumulation:
synchronization, resonance, and infrared (IR) divergence. We have
already studied the cases of synchronization and resonance\cite{Nakamichi2023}.
Here we study the IR divergence for the origin of pink noise in electric
current. 

\section{Dynamical model of electric current }

When we study the electric current, the quantum interference for each
electron does not work properly, as we have discussed in the introduction. 

Then, pink noise in the electric current should be a classical phenomenon.
In semiconductors, the mean free path of an electron is about several
tens of lattice size. Within this size, electrons form a wave packet.
\begin{equation}
\psi(x,t)=\mathrm{e}^{i\left(k_{0}x-\omega_{0}t\right)}\int\phi\left(k_{0}+k^{\prime}\right)\exp\left[ik^{\prime}\left(x-v_{g}t\right)\right]dk^{\prime}\label{eq:1}
\end{equation}
\textcolor{black}{where, $\phi\left(k\right)$ represents the weight
function, and the group velocity $v_{g}=d\omega/dk$ describes the
flow of the wave packet, and the phase can describe the interference.}
An extreme number of wave packets are flowing in a semiconductor sample.
These wave packets interact with impurities in the system and emit
photons. This emission probability of a photon of energy $\hbar\omega$
is proportional to $\omega^{-1}$ and is infrared divergent since
the photon is massless \cite{Itzykson1980}. The wave packets are
exerted by the back reaction of this photon emission and randomly
fluctuate. This dynamics for a single wave packet in the Lagrange
picture (in the coordinate system associated with the wave packet)
will generally be described by the Langevin equation, 
\begin{equation}
\psi''(t)=-\psi(t)+\xi(t)\psi(t)-\kappa\psi'(t)-\lambda\psi(t)^{3},\label{eq:2-1}
\end{equation}
where the random force $\xi$ represents the back reaction of photon
emission with probability $\propto\xi^{-1}$. This product $\xi(t)\psi(t)$
appears here since the interaction is three point form. 

This random field is generated by using the exponential function $\xi=ce^{-\eta}$
where $\eta$ is a uniform random variable, and $c$ is a constant.
Then the distribution function $P(\xi)$ and the $\eta$ distribution
function $p(\eta)\equiv p=const$ are related with each other by $P(\xi)d\xi=p(\eta)d\eta$.
Then,

\begin{equation}
P(\xi)=p(\eta)|d\xi/d\eta|^{-1}=p\xi^{-1}\propto\xi^{-1}.\label{eq:3}
\end{equation}
These frequency-modulated wave packets beat each other and form pink
noise. 

\begin{figure}[tbh]

\includegraphics[width=9cm]{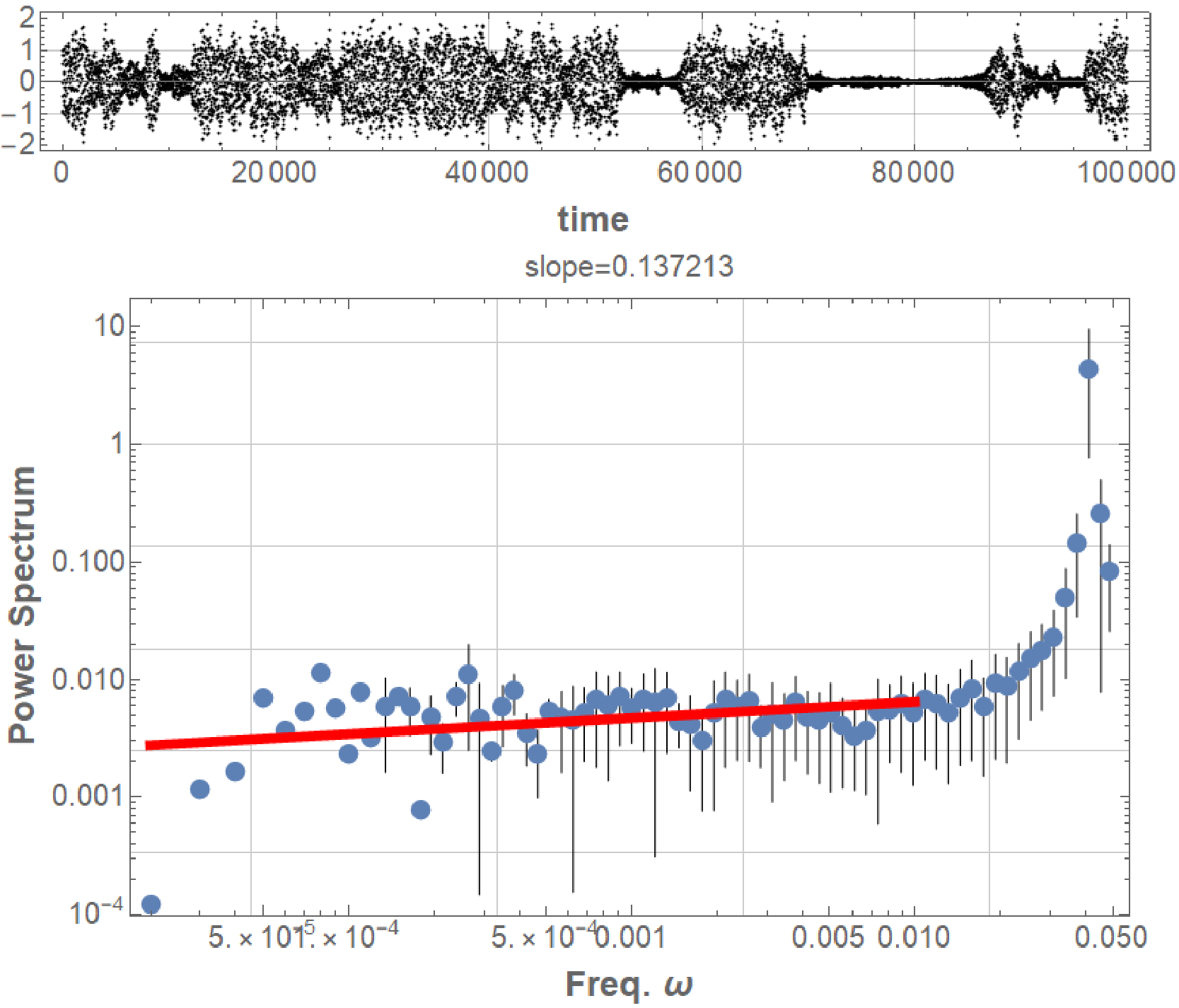}\caption{(upper graph) The solution $\psi(t)$ of Eq.(\ref{eq:2-1}) is drawn.\protect \\
(lower graph) The PSD of it is drawn. This original data does not
show pink noise. We have chosen the parameter $\lambda=0,\kappa=0.001,c=0.9,$
and $\eta$ runs uniformly in the domain $[-2,30].$ The random field
works every time $10^{-2},$ and the total time is $10^{5}$. These
parameters are fixed for all other graphs in this paper. }

\label{fig1}
\end{figure}

As a simple demonstration, we solve Eq.(\ref{eq:2-1}) with appropriate
parameters and calculate the PSD. The PSD of the solution $\psi(t)$
is shown in Fig.\ref{fig1}. As we see, no pink noise appears. 

\begin{figure}[tbh]
\includegraphics[width=9cm]{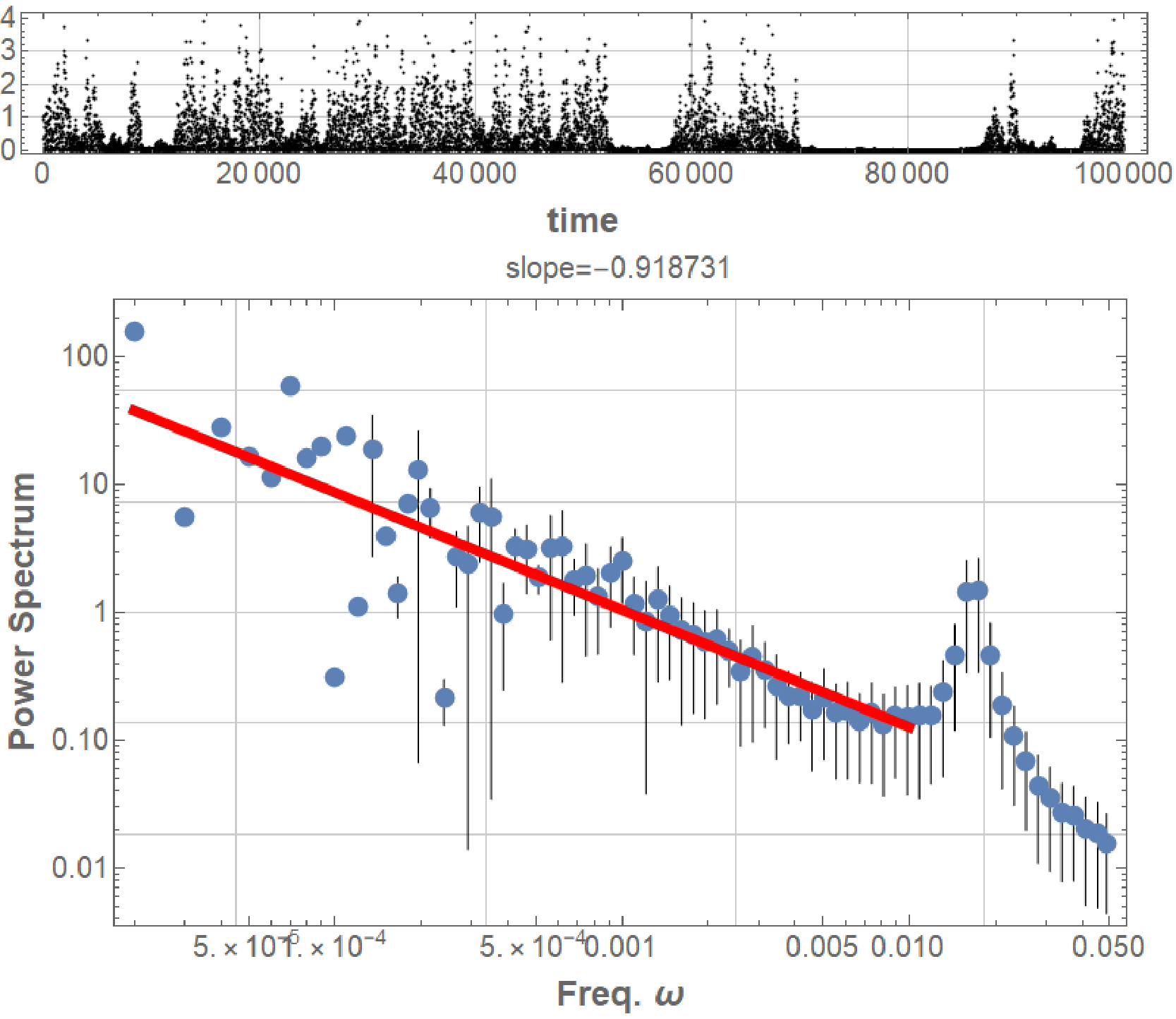}\caption{(upper graph) The time sequence of the \textbf{squared data} $\psi(t)^{2}$.
\protect \\
(lower graph) The PSD of it. This time, the data shows clear pink
noise with a power index of $-0.9.$}
\label{fig2}
\end{figure}

However, the squared data $\psi(t)^{2}$ clearly show pink noise as
in Fig. \ref{fig2}. 

The pink noise is robust, but the choice of parameters alters the
power index. If we increase the self-interaction $\lambda$, the power
index reduces. If we reduce the range of $\eta$, the wave beat is
incomplete, and the range of the power region reduces, while a wider
range of $\eta$ extends the lower law limit in PSD. If we increase
the dissipation $\kappa$, pink noise is dominated by the ordinary
form of PSD in which the power index is $-2$. 

\section{Robust pink noise inheritance}

We want to check the robustness of pink noise. This is because we
expect the pink noise property of the electric current to be inherited
by many other systems and guarantees the variety of pink noise. In
order to do so, we examine the various demodulation processes and
examine the pink noise property. We concretely examine the operations
of the square, absolute values, rectification, thresholding,... Further,
we examine the data after thresholding and then putting all the finite
values 1. This is the timing sequence of the large signal occurrence. 

\begin{figure}[tbh]
\includegraphics[width=9cm]{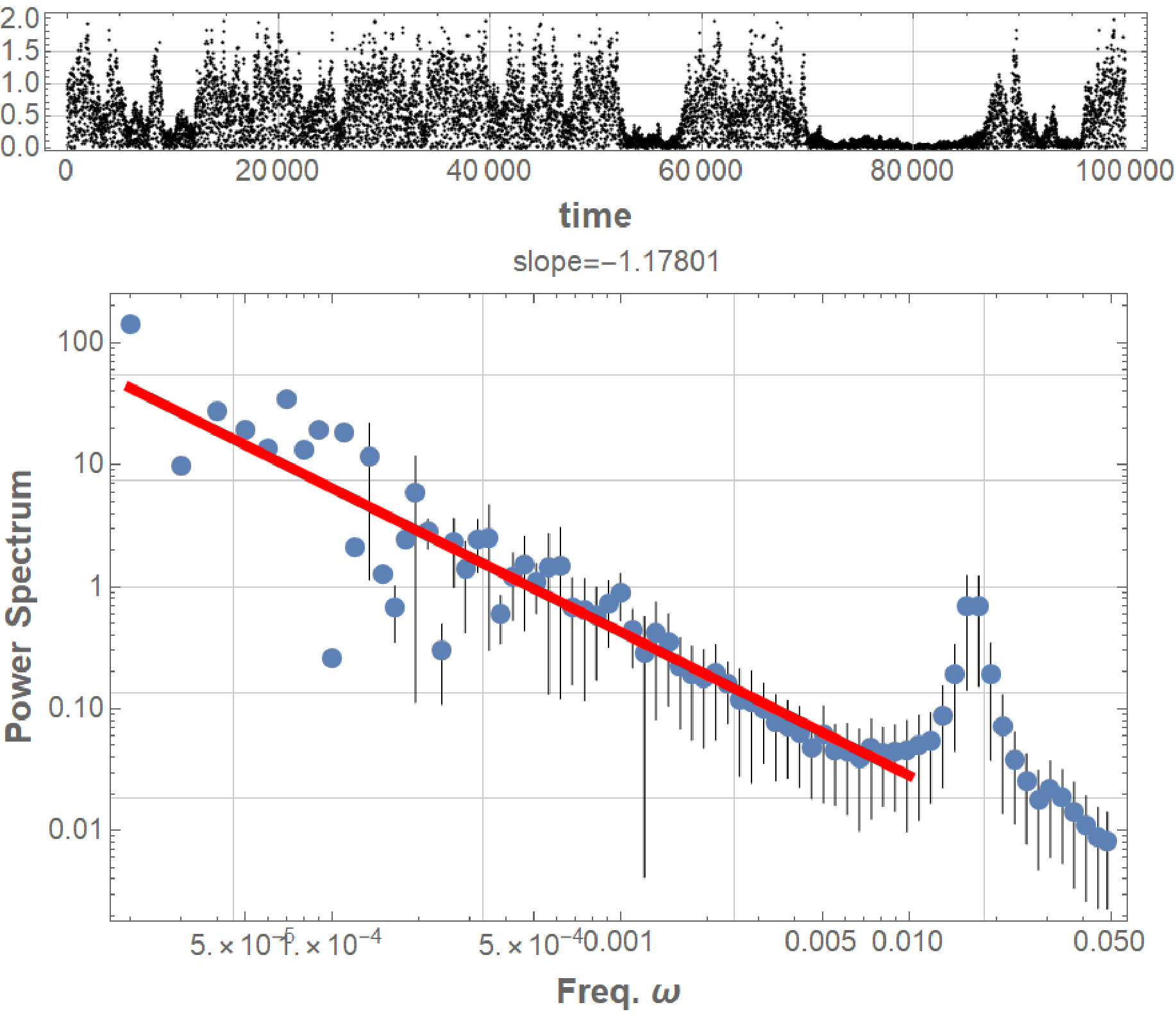}\caption{(upper graph) Time sequence of the \textbf{absolute value} of the
wave packet $\left|\psi(t)\right|$\protect \\
(lower graph) PSD of it, showing pink noise with a power index of
-1.2. }
\label{fig3}
\end{figure}

We first calculate the PSD of the absolute value of the wave packet
$\left|\psi(t)\right|$. The result is in Fig.\ref{fig3}, shows pink
noise as in the case of square, although the power index is slightly
different. Thus, the pink noise is not only associated with loudness. 

\begin{figure}[tbh]
\includegraphics[width=9cm]{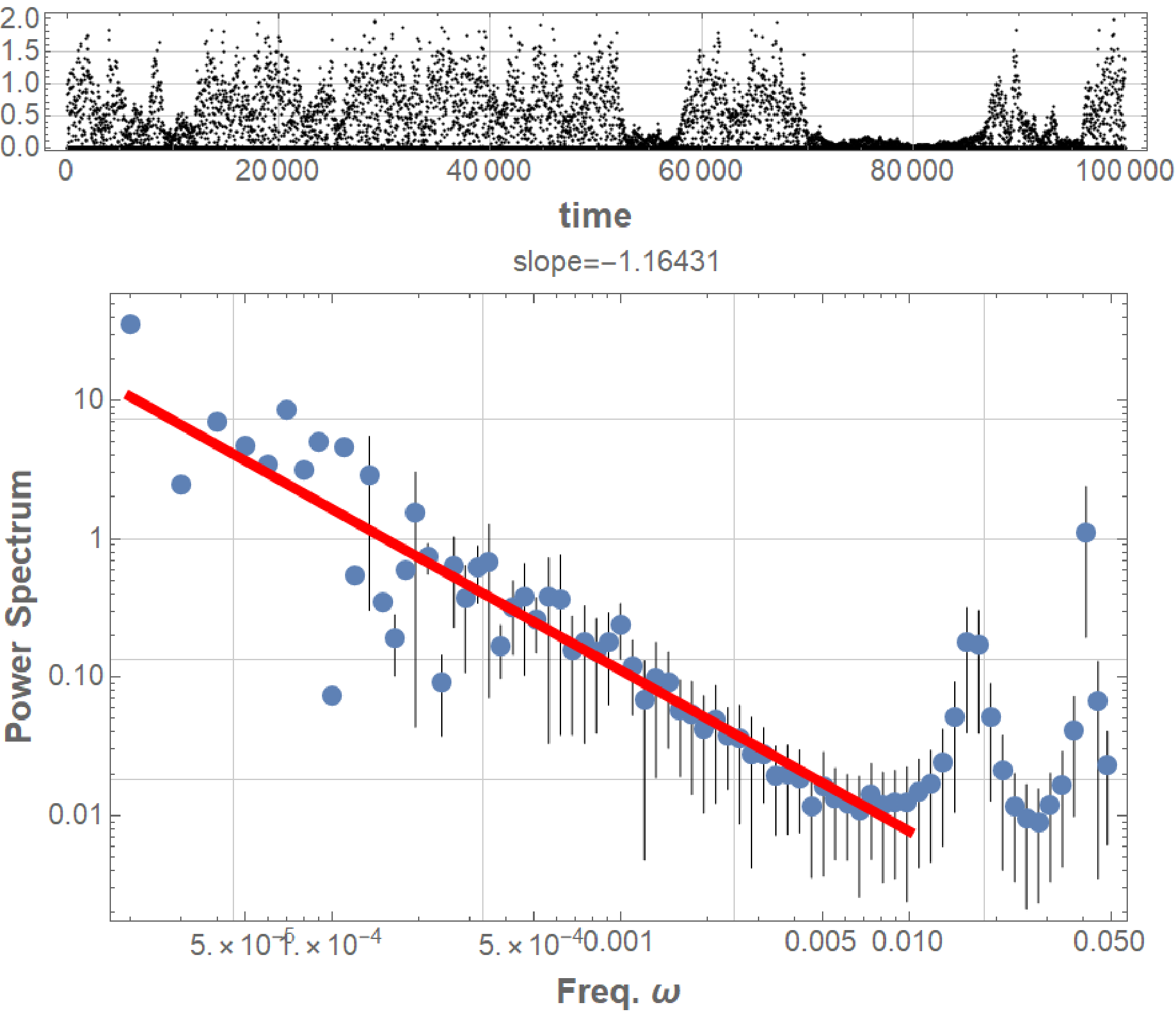}\caption{(upper graph) Time sequence of the \textbf{rectified data} $\psi(t)>0$\protect \\
(lower graph)PSD of it; showing pink noise with a power index of -1.2. }
\label{fig4}
\end{figure}

Similarly, we rectified the wave packet, leaving only a positive signal,
and set the negative signal to zero. Then, this rectified signal shows
pink noise, as in Fig.\ref{fig4}. 

\begin{figure}[tbh]
\includegraphics[width=9cm]{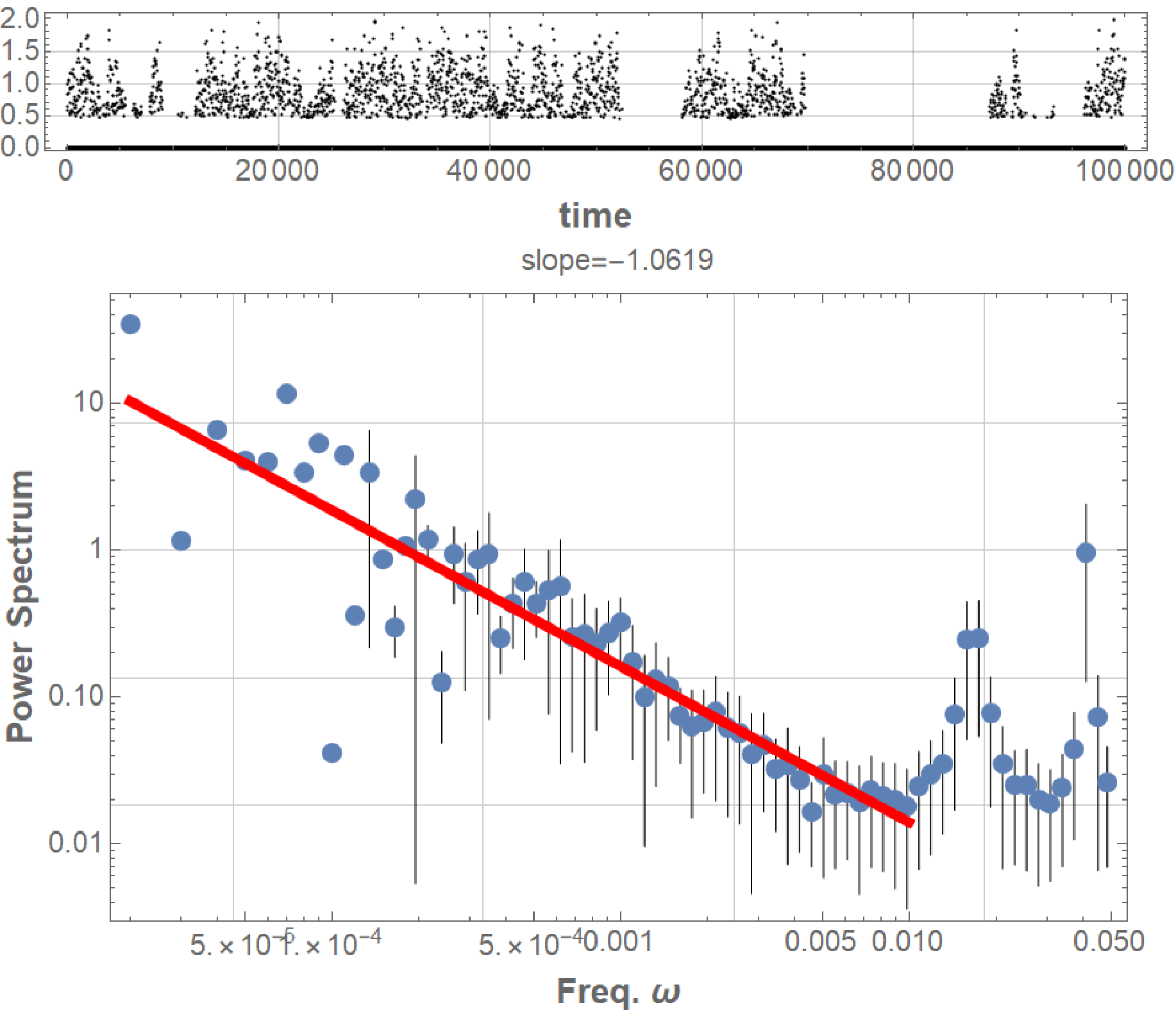}\caption{(upper graph) Time sequence of the \textbf{thresholded data} $\psi(t)>\overline{\left|\psi(t)\right|}$\protect \\
(lower graph)PSD of it; showing pink noise with a power index of -1.1. }
\label{fig5}
\end{figure}

We then tried to set the threshold on the original signal. Leaving
the signal larger than the mean of $\left|\psi(t)\right|$, and set
the rest data to zero. The data still shows pink noise as in Fig.
\ref{fig5}. 

\begin{figure}[tbh]
\includegraphics[width=9cm]{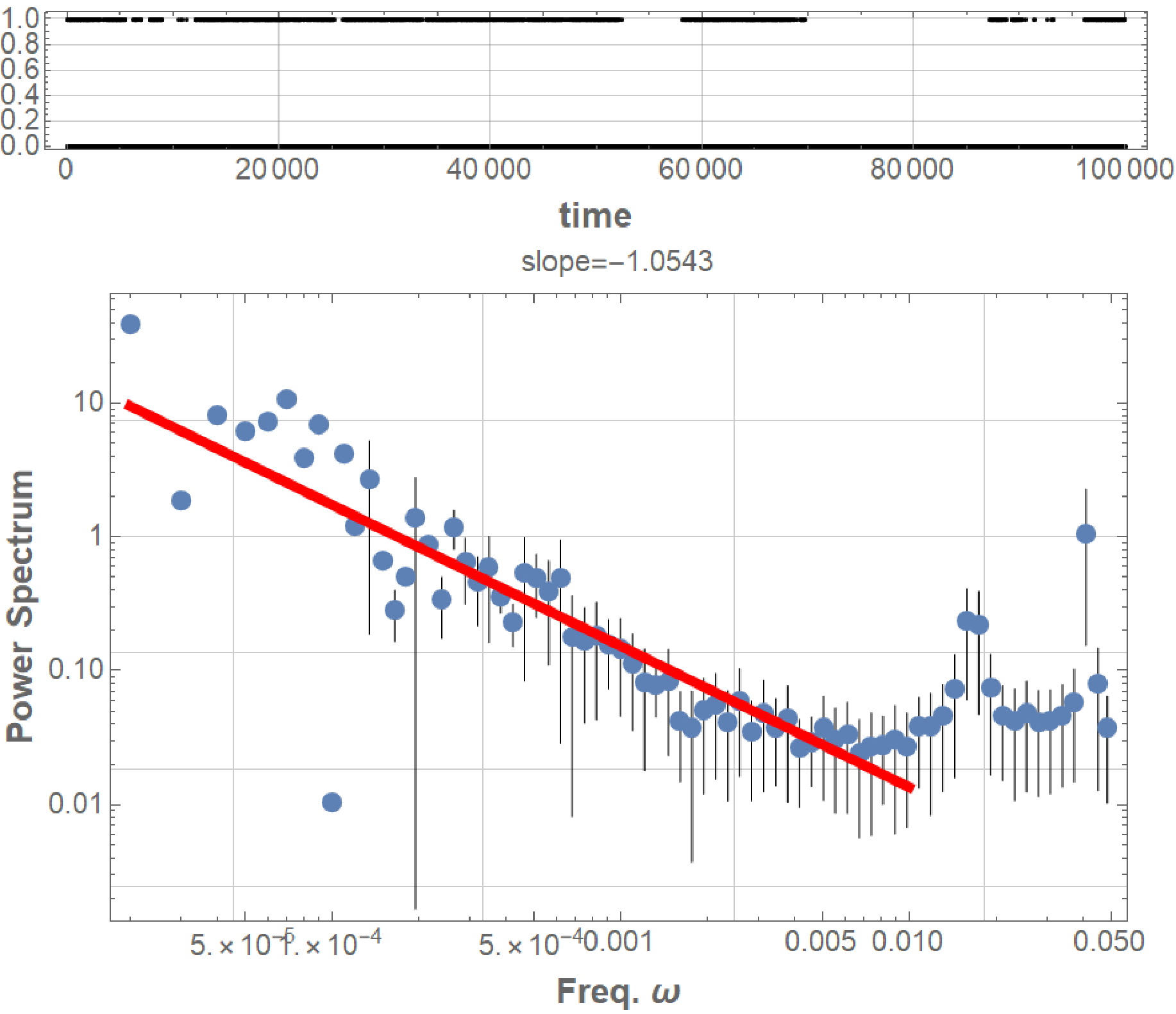}\caption{(upper graph) Time sequence of the \textbf{timing of the thresholed
data}\protect \\
(lower graph)PSD of it, showing pink noise with a power index of -1.1. }
\label{fig6}
\end{figure}

Next, we calculated PSD for the timing of the sequence. We set the
signal larger than the mean of $\left|\psi(t)\right|$ to 1 and set
the rest data to zero. It is remarkable that the PSD still shows clear
pink noise, as in Fig.\ref{fig6}. 

We finally tried the under-threshold data timing. We set the positive
data smaller than the threshold as 1 and set the rest data as zero.
Amazingly, the data still show pink noise, although the power law
range reduces by about one decade. 
\begin{figure}[tbh]
\includegraphics[width=9cm]{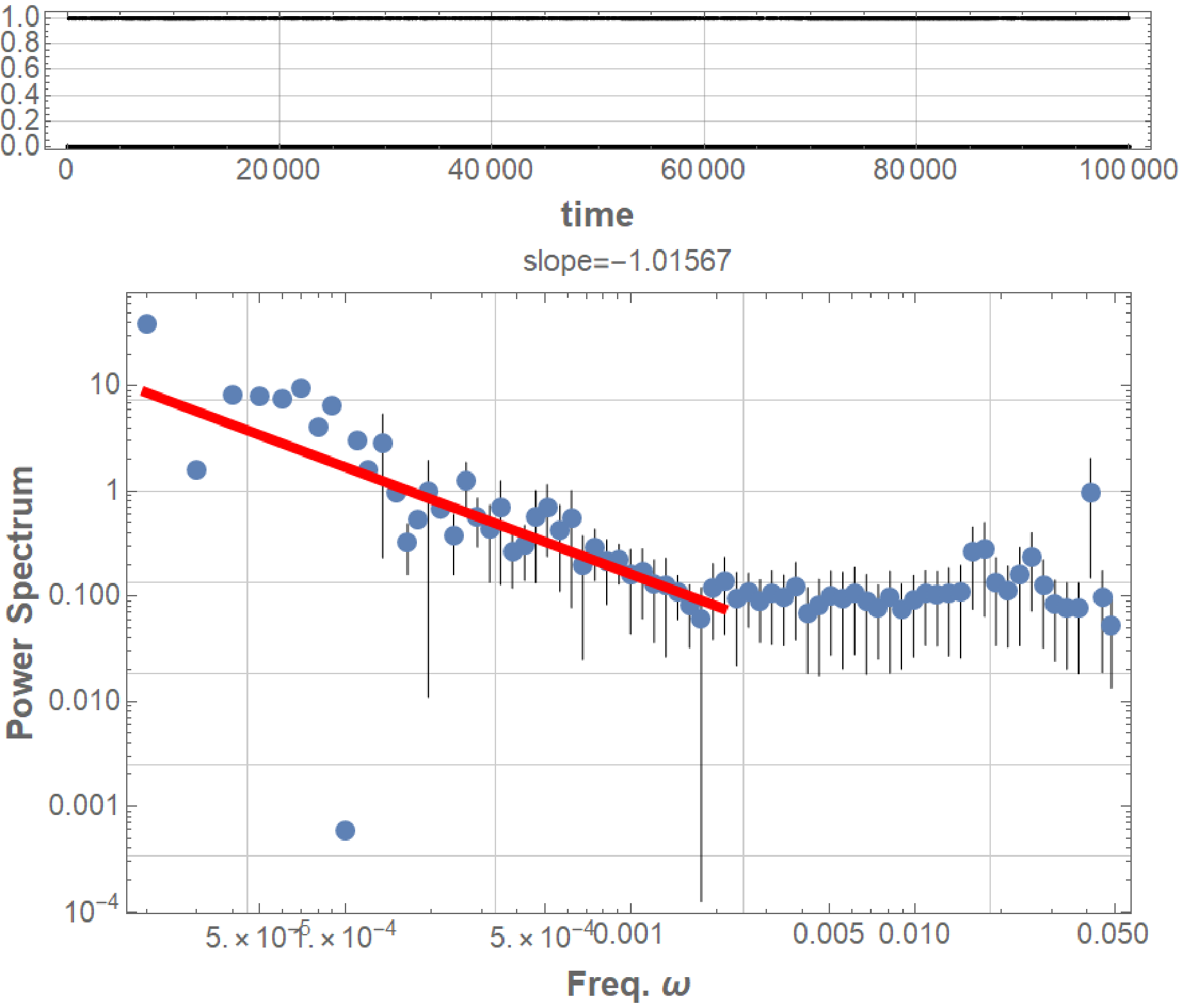}\caption{(upper graph) Time sequence of the \textbf{negative thresholded data}
$0<\psi(t)<\overline{\left|\psi(t)\right|}$\protect \\
(lower graph)PSD of it; showing pink noise with a power index of -1.0.
}
\label{fig7}
\end{figure}

Thus we have demonstrated the robustness of pink noise in the electric
current against various demodulation processes. 

In particular, thresholding (Fig.\ref{fig5}) and timing (Fig. \ref{fig6})
are important to the inheritance from current to the nerve puls system.
Since the electric current governs the living system at the cell level,
the electric current higher than a threshold may yield potential fluctuations
in bio-membranes that go over the threshold of the nerve firing. Although
the model is premature at the present level, we have demonstrated
the possibility that the pink noise is inherited from the current
to the nerve system in the living body. Since the whole process will
be extremely complicated\cite{Musha1997}, we cannot study them here.
However, we have concentrated on the basic property of the inheritance
of the pink noise property by thresholding. 

\section{Possible verifications}

All the above calculations in the previous section show the robustness
of the pink noise in electric current, which will guarantee the inheritance
of the pink noise. They are our predictions for real electric current,
and can be tested in experiments. 

At least in some cases, demodulation procedures are evident. In the
case of pink noise in the vacuum tube \cite{Johnson1925}, Johnson
apparently uses the root mean square operation for the observed voltage
fluctuations in deriving pink noise. This root mean square operation
is equivalent to taking a square and then applying the low-pass filter.
However, we cannot naively use the macroscopic Ohm's law; we cannot
directly compare the voltage and current fluctuations. The electric
current is the square of all the wave packets, while the voltage is
not clear, but probably some integration of the wave packet effects. 

In the case of sound data \cite{Voss1977}, Voss apparently used the
square of the sound wave amplitude, claiming it is the loudness of
the sound. Our proposal is much wider: we can use most of the demodulation
processes as studied in the previous section. 

In the case of the conservative system \cite{Yamaguchi2018}, authors
analyzed the square of the mean field. Our proposal is that if we
take no square operation, the bare mean field never shows pink noise.
Yamaguchi kindly checked this proposal and obtained affirmative results. 

Our model differs from the previous discussions that suppose the multi-fractal
structure or the self-organized criticality. These structures generate
real fluctuations with pink noise. On the other hand, in our model,
the pink noise is apparent and only appears as a beat; the original
signal never shows pink noise. This point should be an adequate criterion
to determine whether the pink noise is real or apparent. 

Our model does not presuppose any elaborated new statistical mechanics
theory. In this sense, we propose a simple and common beat far from
sophisticated physics. We believe this simplicity that makes pink
nise diverse everywhere in nature and in social activities. 

\section{Conclusions}

We have developed an amplitude modulation and demodulation model for
pink noise to the electric current. Further, after demonstrating the
robustness of the pink noise in the model electric current, we proposed
that the pink noise in electric current can inheritate to trigger
the nerve system, thus inducing the pink noise in the living body. 

The above model and proposal are still in the starting stage. We first
want to elaborate our model to include multiple wave packets, 
\begin{equation}
\psi_{total}(t)=\sum_{i=1}^{N}\psi_{i}(t),\label{eq:2}
\end{equation}
where $\psi_{i}(t)$ is the individual wave packet calculated by Eq.(\ref{eq:2-1}).
This total wave packet $\psi_{total}(t)$ must represent a more realistic
electric current in the experiments. We have already partially derived
pink noise in the square $\psi_{total}(t)^{2}$, and this approach
seems to be promising. 

Further, we want to develop a more precise description in the inheritance
from the electric current to the living body activity. There seems
to be much complexity of the constituent electrolytes and possible
bistability structures \cite{Musha1997,Stern1997}. 

We hope we can report our progress soon in our next reports. 

\paragraph*{Acknowledgments}

MM thanks Finnair for providing a sufficient flight connection time
for writing this paper. We thank Chania International Airport Daskalogiannis
and Helsinki-Vantaa Airport for providing a comfortable environment
for our intensive discussions remotely and for completing this paper.

\end{document}